\documentclass[12pt]{article}
\usepackage[T1]{fontenc}
\usepackage{hyperref}
\usepackage{color}
\usepackage[title]{appendix}
\usepackage{authblk}
\usepackage[table]{xcolor} 
\usepackage{amsmath,amssymb,amsfonts,amsthm}
\usepackage[capitalize]{cleveref}
\usepackage{dsfont} 
\usepackage{mathrsfs} 

\newtheorem{modeling}{Modeling}
\newtheorem{problem}{Problem}
\newtheorem{proposition}{Proposition}
\newtheorem{remark}{Remark}
\newtheorem{definition}{Definition}

\newcommand{\eqdef}{\stackrel{\text{def}}{=}}

\renewcommand{\vec}[1]{\ensuremath{\mathbf{#1}}}
\newcommand{\av}{\vec{a}}
\newcommand{\bv}{\vec{b}}
\newcommand{\cv}{\vec{c}}
\newcommand{\ev}{\vec{e}}
\newcommand{\gv}{\vec{g}}
\newcommand{\rv}{\vec{r}}
\newcommand{\xv}{\vec{x}}
\newcommand{\yv}{\vec{y}}

\newcommand{\Am}{\vec{A}}
\newcommand{\Bm}{\vec{B}}
\newcommand{\Cm}{\vec{C}}
\newcommand{\Gm}{\vec{G}}
\newcommand{\Hm}{\vec{H}}
\renewcommand{\Im}{\vec{I}}
\newcommand{\Mm}{\vec{M}}
\newcommand{\Um}{\vec{U}}
\newcommand{\Vm}{\vec{V}}
\newcommand{\Xm}{\vec{X}}
\newcommand{\Ym}{\vec{Y}}
\newcommand{\zerom}{\vec{0}}
\newcommand{\onev}{\mathds{1}}
\newcommand{\tauv}{\vec{\tau}}

\newcommand{\dual}[1]{\ensuremath{{#1}^{\perp}}}
\newcommand{\trsp}[1]{{#1}^{\top}}
\newcommand{\tr}{\ensuremath{\textrm{Tr}}}

\newcommand{\code}[1]{\ensuremath{\mathscr{#1}}}
\newcommand{\Cpub}{{\code{C}_{\mathsf{pub}}}}
\newcommand{\Gpub}{\ensuremath{\Gm_{\mathsf{pub}}}}
\newcommand{\Ginv}{\ensuremath{\Gm_{\mathsf{inv}}}}
\newcommand{\Hpub}{\ensuremath{\Hm_{\mathsf{pub}}}}

\DeclareMathOperator{\Rank}{Rank}
\DeclareMathOperator{\Diag}{Diag}
\newcommand{\minor}[2]{\left\vert #1 \right\vert_{#2}}

\begin{document}

\title{Improvement of algebraic attacks for solving superdetermined MinRank instances}
	
\author[1]{Magali Bardet \thanks{magali.bardet@univ-rouen.fr}}
\author[1]{Manon Bertin \thanks{manon.bertin@etu.univ-rouen.fr}}

\affil[1]{LITIS, University of Rouen Normandie, France.}

\maketitle

\begin{abstract}
  The MinRank (MR) problem is a computational problem that arises in
  many cryptographic applications. In Verbel et al.~\cite{VBCPS19},
  the authors introduced a new way to solve superdetermined instances
  of the MinRank problem, starting from the bilinear Kipnis-Shamir
  (KS) modeling. They use linear algebra on specific Macaulay
  matrices, considering only multiples of the initial equations by one
  block of variables, the so called  ``kernel'' variables. Later, Bardet
  et al.~\cite{BBCGPSTV20} introduced a new Support Minors modeling
  (SM), that consider the Plücker coordinates associated to the kernel
  variables, {\em i.e.} the maximal minors of the Kernel matrix in the KS modeling.

  In this paper, we give a complete algebraic explanation of the link
  between the (KS) and (SM) modelings (for any instance).  We then
  show that superdetermined MinRank instances can be seen as easy
  instances of the SM modeling. In particular, we show that performing
  computation at the smallest possible degree (the ``first degree
  fall'') and the smallest possible number of variables is not always
  the best strategy. We give complexity estimates of the attack for
  generic random instances.

  We apply those results to the DAGS cryptosystem, that was submitted
  to the first round of the NIST standardization process. We show that
  the algebraic attack from Barelli and Couvreur~\cite{BC18}, improved
  in Bardet et al.~\cite{BBCO19}, is a particular superdetermined
  MinRank instance. Here, the instances are not generic, but we show
  that it is possible to analyse the particular instances from DAGS
  and provide a way to select the optimal parameters (number of
  shortened positions) to solve a particular instance.

  \paragraph{keywords} {Post-quantum cryptography
-- MinRank problem 
  -- algebraic attack -- DAGS cryptosystem.}
\end{abstract}
	
\section{Introduction}
\subsubsection*{The MinRank Problem}
The MinRank problem was first mentioned in~\cite{BFS99} where its
NP-completeness was also proven. It is a central problem in algebraic
cryptanalysis, starting with the Kipnis and Shamir
modeling~\cite{KS99} for the HFE encryption scheme. The MinRank
problem is very simple to state:

\begin{problem}[Homogeneous MinRank problem]~\\
  \indent\emph{Input}: a target rank $r \in \mathbb N$ and $K$ 
  matrices $\Mm_{1},\dots,\Mm_{K} \in \mathbb{F}_{q}^{m\times n}$.\\
  \indent\emph{Output}: field elements $x_1,x_2,\dots,x_K \in \mathbb{F}_{q}$, not all zero, such that
  \begin{equation*}
    \Rank\left(\Mm_{\xv}\eqdef \sum_{i=1}^{K}x_i\Mm_{i} \right) \leqslant r.
  \end{equation*}
\end{problem}
It plays a central role in public key cryptography. Many multivariate
schemes are strongly related to the hardness of this problem, as
in~\cite{P96,KS99,PCYTD15,PBD14}. The 3rd round NIST post-quantum
competition finalist Rainbow~\cite{DCPSY19}, or alternate
GeMSS~\cite{CFMPPR19} suffered attacks based on the MinRank
problem~\cite{B21,B22,TPD21,BBCPSV21}.

In code-based cryptography, the MinRank problem is exactly the
decoding problem for matrix codes in rank metric. The two submissions
ROLLO and RQC~\cite{ABDGHRTZABBBO19,AABBBBCDGHZ20}, from the 2nd round
of the NIST post-quantum competition, have been attacked using algebraic
cryptanalysis in~\cite{BBBGNRT20,BBCGPSTV20}. Their security analysis 
relies on the decoding problem for $\mathbb{F}_{q^m}$-linear rank-metric codes, which 
can actually be reduced to the MinRank problem. 

This is of great importance for cryptographic purposes to design
algorithms that solve efficiently algebraic modeling for the MinRank
problem, and to understand their complexity.

\subsubsection*{Algebraic modeling}
There has been a lot of recent progress in the algebraic modeling and
solving of the MinRank problem. We start by recalling the first
modeling, namely the Kipnis-Shamir (KS) modeling.  Note that it
implicitely assumes that the $n-r$ first columns of the small-rank
matrix $\Mm_{\xv}$ we are looking for are linearly dependent from the
last $r$ ones. In this paper, we will assume that we are looking for a
matrix $\Mm_{\xv}$ of rank \emph{exactly} $r$ (this can be achieved
by looking for increasing ranks, starting from $r=1$), and that the
last $r$ columns of $\Mm_{\xv}$ are linearly independent (it is true
up to a permutation of the columns, and for random matrices it is true
with high probability). We will see later that this last assumption is
not mandatory.

\begin{modeling}[Kipnis-Shamir Modeling~\cite{KS99}]
  Consider a MinRank instance
  $(\Mm_{1},\dots,\Mm_{K})\in \mathbb{F}_{q}^{m\times n}$ with target rank
  $r$. Then, the MR problem can be solved by finding
  $x_1,\dots,x_K\in \mathbb{F}_{q}^K$, and
  $\Cm = (c_{i,j})\in \mathbb{F}_{q}^{r\times (n-r)}$ such that
  \begin{align}\label{eq:KS}
    \left(\sum_{i=1}^K x_i \Mm_{i}\right)
    \begin{pmatrix}
      \Im_{n-r}\\{\Cm}
    \end{pmatrix} = \zerom_{m\times (n-r)}.\tag{KS}
  \end{align}
  The $m(n-r)$ equations are bilinear in the $K$ \emph{linear
    variables} $\xv = (x_1,\dots,x_K)$ and the $r(n-r)$ entries of the
  formal matrix $\Cm = (c_{i,j})_{i,j}$, refered to as the
  \emph{kernel variables}.
\end{modeling}
It is clear that a matrix has rank $\le r$ if and only if its right
kernel has dimension at least $n-r$, so that any solution of the
MinRank problem is a solution of~\eqref{eq:KS}.

The complexity of solving a generic bilinear system has been studied
in \cite{FSS10,FSS11}, and gives an upper bound for the KS system, but
this estimate wildly overestimates the experimental results. 

The matrix $\Mm_{\xv}$ has rank $\le r$ if and only if all its minors
of size $r+1$ are zero. This modeling has been presented and analysed
in~\cite{FSS13,FSS10}.  Under the assumption that the last $r$ columns
of $\Mm_{\xv}$ are linearly independent, it is sufficient to consider minors involving columns in
sets $T = \lbrace t\rbrace \cup \lbrace n-r+1..n\rbrace$ with
$1\le t \le n-r$, as it means that the last $r$ columns generate the
column vector space.
The notation $\minor{\Mm_{}}{J,T}$ represents the determinant of the submatrix of $\Mm$ where we keep only rows in $J$ and columns in $T$.
\begin{modeling}[Minors Modeling]
  Let $(\Mm_{1},\dots,\Mm_{K})\in \mathbb{F}_{q}^{m\times n}$ be a MinRank
  instance with target rank $r$. Then, the MR problem can be solved by
  finding $x_1,\dots,x_K\in \mathbb{F}_{q}^K$ such that
  \begin{align}\label{eq:minors}
    \left\lbrace    \minor{\Mm_{\xv}}{J,T}
    = 0, \forall J\subset\lbrace 1..m\rbrace, \#J=r+1, T=\lbrace t\rbrace \cup \lbrace n-r+1..n\rbrace\subset\lbrace 1..n\rbrace \right\rbrace\tag{Minors}
  \end{align}  
\end{modeling}

Recently, a new modeling has been introduced in~\cite{BBCGPSTV20},
that is at the moment the most efficient one from the complexity point
of view. It uses two ideas, that we will separate in two modelings: the first idea is that~\eqref{eq:KS} means that the vector space with generator
matrix $\Mm_{\xv}$ is orthogonal to the one with generator matrix $
\begin{pmatrix}
  \Im_{n-r} & \trsp{\Cm}
\end{pmatrix}$. It is then straightforward to see that any row of $\Mm_{\xv}{}$ belongs to the dual space
with generator matrix $\begin{pmatrix} -{\Cm} & \Im_{r}
\end{pmatrix}$. This leads to:
\begin{modeling}[Support Minor Modeling-$\Cm$ \cite{BBCGPSTV20}]
Let $(\Mm_{1},\dots,\Mm_{K})\in \mathbb{F}_{q}^{m\times n}$ be a MinRank
  instance with target rank $r$. Then, the MR problem can be solved by
  finding $x_1,\dots,x_K\in \mathbb{F}_{q}^K$, and
  $\Cm = (c_{i,j})_{1\le i \le r, 1\le j \le n-r}\in \mathbb{F}_{q}^{r\times
    (n-r)}$ such that
  \begin{align}\label{eq:SMM}
    \left\lbrace    \minor{\begin{pmatrix}
          \rv_i\\
          -{\Cm} \;  \Im_{r} 
        \end{pmatrix}}{*,T} = 0, \; \forall T\subset\lbrace
      1..n\rbrace, \#T=r+1, \text{ and } \rv_i \text{ row of }
      \Mm_{\xv}{} \right\rbrace.\tag{SM-\Cm}
  \end{align}
  The $m\binom{n}{r+1}$ equations are bi-homogeneous with bi-degree $(1,r)$ in the $K$ \emph{linear
    variables} $\xv = (x_1,\dots,x_K)$ and the $r(n-r)$ entries of the
  formal matrix $\Cm = (c_{i,j})_{i,j}$, refered to as the
  \emph{kernel variables}. 
\end{modeling}
Note that in~\eqref{eq:SMM}, the entries of $\Cm$ appear only as
maximal minors of $(-\Cm \; \Im_r)$. This leads to the second idea
from~\cite{BBCGPSTV20}, which consists in using the Plücker
coordinates: we replace each $\minor{(-\Cm\; \Im_r)}{*,T}$, that is a
polynomial of degree $\#T=r$ in the entries of $\Cm$, by a new
variable $c_T$ using the injective Plücker map, see~\cite[p.6]{BV88}. 
\begin{align*}
  p :  \lbrace \mathcal W \subset  \mathbb{F}_{q}^n : \dim(\mathcal W)=r\rbrace &\to \mathbb P^N(\mathbb{F}_{q}) & (N=\textstyle{\binom{n}{r}}-1)\\
  \Cm & \mapsto  (c_T)_{T\subset\lbrace 1..n\rbrace, \#T = r}.
\end{align*}

\begin{modeling}[Support Minor Modeling-$c_T$\cite{BBCGPSTV20}]
  Let $(\Mm_{1},\dots,\Mm_{K})\in \mathbb{F}_{q}^{m\times n}$ be a MinRank
  instance with target rank $r$. Then, the MR problem can be solved by
  finding $x_1,\dots,x_K\in \mathbb{F}_{q}^K$, and $(c_T)_{T\subset\lbrace 1..n\rbrace, \#T=r}\subset \mathbb{F}_{q}^{\binom{n}{r}}$
  such that
  \begin{align}\label{eq:SMMcT}
    \left\lbrace
    \sum_{t\in T} (\Mm_{\xv}{})_{i,t} c_{T\setminus\lbrace t\rbrace}=0, \; \forall T\subset\lbrace
      1..n\rbrace, \#T=r+1, \text{ and } i\in\lbrace 1..m\rbrace \right\rbrace.\tag{SM}
  \end{align}
  The $m\binom{n}{r+1}$ equations are bilinear in the $K$ \emph{linear
    variables} $\xv = (x_1,\dots,x_K)$ and the $\binom{n}{r}$
  \emph{minor variables} $c_{T}$, 
  for all $T\subset\lbrace 1..n\rbrace, \#T=r$.
\end{modeling}
The benefit of introducing such coordinates to describe a vector space
rather than a matrix describing a basis is that contrarily to the
matrix representation, a vector space $W$ has unique Pl\"ucker
coordinates associated to it. This is not the case of the matrix
representation of a vector space: if the rows of a matrix $\Cm$
generate the vector space, then the rows of $\Am\Cm$ generate the same
vector space for any invertible $\Am\in GL(r,\mathbb{F}_{q})$. For our algebraic
system, this brings the benefit of reducing the number of solutions of
the system: there are several solutions $\Cm$ to the algebraic system
\eqref{eq:SMM}, that correspond to one unique solution
to~\eqref{eq:SMMcT}.  As already pointed out in \cite{BBCGPSTV20}, it
is also extremely beneficial for the computation to replace
polynomials $\minor{(-\Cm\;\Im_r}{*,T}$ with $r!$ terms of degree
$r$ in the entries of $\Cm$ by single variables $c_T$'s in $\mathbb{F}_{q}$.  We
will use~\eqref{eq:SMM} for the theoretical analysis of the link
between the various modelings, and~\eqref{eq:SMMcT} for the
computational solving.

\subsubsection*{Contributions}
As a first contribution, we show that the first three systems are related, more precisely
\begin{proposition}
  \label{prop:relations}
  The set of equations~\eqref{eq:KS} is included in the set of
  equations \eqref{eq:SMM}, and the ideals generated by the
  \ref{eq:SMM} and \ref{eq:KS} equations are equal.
  Equations~\ref{eq:minors} (Minors modeling) 
  are included in the ideal generated by~\ref{eq:KS}. 
\end{proposition}
This proposition applies to any instance, without particular
hypothesis.

Note that~\cref{eq:minors} contain only the linear variables, hence
the ideals cannot be equal.

This proposition is not only interesting on the theoretical point of
view, but it also allows to understand different computational
strategies and to select the best one. A discussion is provided
in~\cref{remark:strategy}.

In~\cite{VBCPS19}, Verbel et al. analyse degree falls occuring during
a Gröbner basis computation of~\eqref{eq:KS}, and show that for
overdetermined systems this can occur before degree $r+2$, which is
the general case. As a second contribution, we show that these degree
falls are in fact equations from~\ref{eq:SMM}. Using the Plücker
coordinates in~\eqref{eq:SMMcT} allows to drastically reduce the size
of the considered matrices. Moreover, we give example to show that
minimising the degree and number of equations is not always the best
strategy for optimising the solving complexity.

Finally, we revisit the DAGS cryptosystem~\cite{BBBCDGGHKNNPR17}, that
was a 1st round candidate to the NIST post-quantum standardization
process, and was attacked by Barelli and Couvreur~\cite{BC18}. We show
that the attack is in fact a MinRank attack, and describe the
structure of this non-random superdetermined MinRank instance.  This
precise understanding of the problem makes it possible to choose the
right parameters for an optimal attack.

\section{Notation and preliminaries}
Vectors are denoted by lower case boldface letters such as $\xv,~\ev$
and matrices by upper case letters $\Cm,~ \Mm$. The all-zero vector of
length $\ell$ is denoted by $\zerom_\ell$. The $j$-th coordinate of a
vector $\xv$ is denoted by $x_{j}$ and the submatrix of a matrix $\Cm$
formed from the rows in $I$ and columns in $J$ by $\Cm_{I,J}$. When
$I$ (\textit{resp.}\; $J$) represents all the rows (\textit{resp.}\; columns), we may use
the notation $\Cm_{*, J}$ (\textit{resp.}\; $\Cm_{I,*}$). We simplify
$\Cm_{i,*} = \Cm_{\lbrace i\rbrace, *}$ (\textit{resp.}\;
$\Cm_{*,j} = \Cm_{*,\lbrace j \rbrace}$) for the $i$-th row of
$\Cm$ (\textit{resp.}\; $j$-th column of $\Cm$) and
$c_{i,j}= \Cm_{\lbrace i\rbrace, \lbrace j\rbrace} $ for the entry
in row $i$ and column $j$. Finally, $\minor{\Cm}{}$ is
the determinant of a matrix $\Cm$, $\minor{\Cm}{I,J}$
is the determinant of the submatrix $\Cm_{I,J}$ and
$\minor{\Cm}{*,J}$ the one of $\Cm_{*, J}$. The transpose of a matrix
$\Cm$ is $\trsp{\Cm}$.

The all-one vector of size $n$ is denoted by $\onev_n = (1,\dots,1)$.

To simplify the presentation, we restrict ourselves to a field of
characteristic 2, but the results are valid for any characteristic,
the only difference being the occurrence of a $\pm$ sign before each
formula. 

For any matrix $\Am$ of size $q\times r$ with $r\le q$, and any set
$J\subset \lbrace 1..q\rbrace $ of size $r+1$, we define the vector $ \Vm_J(\Am)$ of length $q$ whose $j$st entry is 0 if $j\notin J$ and $\minor{\Am}{J\setminus\lbrace j \rbrace,*}$ for $j\in J$. For $\Am$ of size $r\times q$ with $r\le q$ we define $\Vm_J(\Am)\eqdef \Vm_J(\trsp{\Am})$.

Using Laplace expansion along a column, it is clear that for any vector $\av$ of length $q$ we have
\begin{align}
  \Vm_J(\Am) \trsp{\av} &= \minor{
                   \begin{pmatrix}
                     \trsp{\av}& \Am
                   \end{pmatrix}}{J,*}.\label{eq:Vm}
\end{align}

We denote by $ vec_{row}(\Am)$ (\textit{resp.}\; $vec_{col}(\Am)$) the vertical
vector formed by concatenating successives rows (\textit{resp.}\; cols) of $\Am$. We have the formula
\begin{align}
  \label{eq:formulavecrow}
  vec_{row}(\Am\Xm\Ym) &= (\Am\otimes \trsp{\Ym})vec_{row}(\Xm)\\
  vec_{col}(\Am\Xm\Ym) &= (\trsp{\Ym}\otimes {\Am})vec_{col}(\Xm)\notag{}
\end{align}
where $\Am\otimes \Bm \eqdef (a_{i,j}\Bm)_{i,j}$ is the Kronecker product of two matrices $\Am=(a_{i,j})_{i,j}$ and $\Bm$.

For a system $\mathcal F=\lbrace f_1,\dots,f_M\rbrace$ of bilinear
equations in two sets of variables $\xv=(x_j)_{1\le j\le n_x}$ and
$\yv=(y_\ell)_{1\le \ell\le n_y}$, it is usual to consider the
associated Jacobian matrices:
\begin{align*}
  Jac_{\xv}(\mathcal F)
  &=
    \begin{pmatrix}
      \frac{\partial f_i}{\partial x_j}
    \end{pmatrix}_{i=1..M, j=1..n_x},
  &
    Jac_{\yv}(\mathcal F)
  &=
    \begin{pmatrix}
      \frac{\partial f_i}{\partial y_\ell}
    \end{pmatrix}_{i=1..M, \ell=1..n_y}.
\end{align*}
For homogeneous bilinear systems they satisfy the particular relation:
\begin{align*}
  Jac_{\xv}(\mathcal F)\trsp{\xv} &= \trsp{
                                    \begin{pmatrix}
                                      f_1 & \dots & f_M
                                    \end{pmatrix}}
\end{align*}
and any vector in the left kernel of a Jacobian matrix is a syzygy of
the system.  Moreover, $Jac_{\xv}$ is a matrix whose entries are
linear form in the variables $\yv$, and Cramer's rule show that the
left kernel of $Jac_{\xv}$ contains vectors $\Vm_T(Jac_{\xv}(\mathcal F)_{T,*})$ using the notation from~\eqref{eq:Vm}
for all $T\subset \lbrace 1..M\rbrace$ of size $n_y+1$. 
Generically those vectors generate the left kernel.  For affine
systems, we consider the jacobian matrix associated to the homogeneous
part of highest degree of the system, and any syzygy for this part,
that is not a syzygy of the entire system leads to a degree fall.

\section{Relations between the various modelings}
This section applies to any MinRank instance without any specific hypothesis.

The KS modeling consists in bilinear equations in two blocks of
variables $\xv$ and $\Cm$, whereas the \ref{eq:SMM} modeling contains equations
of degree $1$ in $\xv$ and $r$ in $\Cm$, the variables $\Cm$ appearing
only as maximal minors of $\textstyle{
  \begin{pmatrix}
    -{\Cm} & \Im_r
  \end{pmatrix}}$.

In the case of the KS modeling, it has been noticed in~\cite{VBCPS19},
and later~\cite[Lemma1]{BBCGPSTV20} that the jacobian matrices have a
very particular shape: if we write
$\textstyle{\Mm_{\xv}{} =
\begin{pmatrix}
  \Mm_{\xv}^{(1)} & \Mm_{\xv}^{(2)}
\end{pmatrix}}$ with $\Mm_{\xv}^{(1)}$ of size $m\times (n-r)$ and
$\Mm_{\xv}^{(2)}$ of size $m\times r$, and in the same way we write each
$\Mm_{i}{} =
\begin{pmatrix}
  \Mm_{i}^{(1)} & \Mm_{i}^{(2)}
\end{pmatrix}$, then the homogeneous part of highest degree of the
system is $\Mm_{\xv}^{(2)}{\Cm}$, and its Jacobian matrices are, if we take
the variables and equations in row/column order:
\begin{align}
  &Jac_{x_i}\left(vec_{row}\left(x_i\Mm_{i}^{(2)}{\Cm}\right)\right)
   = vec_{row}\left(\Mm_{i}^{(2)}{\Cm}\right) = \left(\Im_{m}\otimes{\trsp{\Cm}}\right)
       vec_{row}\left(\Mm_{i}^{(2)}\right)
\notag{}\\
 &  Jac_{\xv}\left(vec_{row}\left(\Mm_{\xv}^{(2)}{\Cm}\right)\right)
   = \left(\Im_{m}\otimes{\trsp{\Cm}}\right)
 \begin{pmatrix}
       vec_{row}\left(\Mm_1^{(2)}\right) & \dots & vec_{row}\left(\Mm_{K}^{(2)}\right)
     \end{pmatrix}\notag{}\\
&  Jac_{vec_{col}(\Cm)}\left(vec_{col}\left( \Mm_{\xv}^{(2)} {\Cm}
  \right)\right) = \Im_{n-r} \otimes \Mm_{\xv}^{(2)}\label{eq:jacCm}
\end{align}
The jacobian matrix in $\Cm$ admits a left kernel that contains the following vectors:
  \begin{align}
    \ev_i \otimes \Vm_J(\Mm_{\xv}^{(2)}) & \text{ for any } J\subset\lbrace 1..p\rbrace, \#J = r+1, 1\le i \le n-r,
  \end{align}
  where $\ev_i$ is the $i$th row of $\Im_{n-r}$. As a consequence, the
  ideal generated by the~\eqref{eq:KS} equations contains the
  equations
 \begin{align*}
   ( \ev_i \otimes \Vm_J(\Mm_{\xv}^{(2)})) vec_{col}(\Mm_{\xv}{}
   \begin{pmatrix}
     \Im_{n-r}\\{\Cm}
   \end{pmatrix}) &= 
                    ( \ev_i \otimes \Vm_J(\Mm_{\xv}^{(2)})) vec_{col}(\Mm_{\xv}{1})\\
                  &=  \Vm_J(\Mm_{\xv}^{(2)})\Mm_{\xv}^{(1)}\trsp{\ev_i} (\text{  thanks to~\eqref{eq:formulavecrow}})\\
   &= \minor{\Mm_{\xv}{}}{J,\lbrace i\rbrace \cup\lbrace n-r+1..n\rbrace} (\text{ thanks to~\eqref{eq:Vm}}).
 \end{align*}
 Those are precisely the~\eqref{eq:minors} equations.
 
 The jacobian matrix in $\xv$ admits a left kernel that contains the vectors
 \begin{align}
   \ev_\ell\otimes \Vm_J({\Cm})&\text{ for any } J\subset\lbrace 1..n-r\rbrace, \#J=r+1, 1\le \ell \le m.
 \end{align}
 The ideal generated by the~\eqref{eq:KS} equations contains the equations
 \begin{align*}
   (\ev_\ell\otimes \Vm_J({\Cm})) vec_{row}(\Mm_{\xv}^{(1)}) &= \ev_\ell \Mm_{\xv}^{(1)} \trsp{\Vm_J({\Cm})}  =  \Vm_J({\Cm})\trsp{({\Mm_{\xv}^{(1)}}_{\ell,*})}\\
   &= \minor{
     \begin{pmatrix}
       \trsp{({\Mm_{\xv}^{(1)}}_{\ell,*})} & \trsp{\Cm}
     \end{pmatrix}}{J,*} =\minor{
     \begin{pmatrix}
{\Mm_{\xv}^{(1)}}_{\ell,*} \\ {\Cm}
     \end{pmatrix}}{*,J}.
 \end{align*}
 They are exactly the~\eqref{eq:SMM} equations for
 $J\subset\lbrace 1..n-r\rbrace$. They have a degree $r$ in the kernel variables $c_{i,j}$.
 
In~\cite{VBCPS19}, the authors propose to solve (any)
instances of KS by considering particular elements in the left kernel
of the Jacobian matrix in $\xv$ for some degree $1\le d\le r-1$. This is
done by considering all combination of the polynomials with coefficients
\begin{align}
\ev_{\ell}\otimes \Vm_J({\Cm_{T,*}})
\end{align}
for any $d\in\lbrace 1..r\rbrace$, $J\subset\lbrace 1..n-r\rbrace$, $\#J=d+1$,
$T\subset\lbrace 1..r\rbrace$, $\#T=d$ and $\ell\in \lbrace 1..m\rbrace$. The authors in~\cite[Theorem
2]{VBCPS19} construct a matrix $\Bm_J$ whose left kernel contains
elements related to the left kernel of the Jacobian matrix in
$\xv$. The key remark is that the equations they consider are
\begin{align*}
(  \ev_{\ell}\otimes \Vm_J({\Cm_{T,*})}) vec_{row}(\Mm_{\xv}
  \begin{pmatrix}
    \Im_{n-r}\\{\Cm}
  \end{pmatrix}) &={\ev_{\ell}} \Mm_{\xv}  \begin{pmatrix}
    \Im_{n-r}\\{\Cm}
  \end{pmatrix} \trsp{\Vm_J({\Cm_{T,*}})}\\
		 &= \minor{
    \begin{pmatrix}
      \rv_\ell\\
      -{\Cm} \; \Im_{r}
    \end{pmatrix}}{*,T'}
\end{align*}
for
$T'=J \cup (\lbrace n-r+1..n\rbrace\setminus (T+n-r))\subset\lbrace
1..n\rbrace$ of size $r+1$.  The equations have indeed a degree $d$ in the kernel variables $c_{i,j}$.

Note that for $d=0$, for
$T'=\lbrace \ell\rbrace \cup \lbrace n-r+1..n\rbrace$ the equation $\minor{
  \begin{pmatrix}
\rv_\ell\\-{\Cm} \; \Im_r
\end{pmatrix}}{*,T'}$ is the
$\ell$th KS equation, and we get all SM
equations.  As a consequence, we have proven~\cref{prop:relations}.

\begin{remark}
  \label{remark:strategy}
  In the light of the previous results, we can understand more
  precisely the behavior of a generic Gröbner basis (GB) algorithm
  with a graded monomial ordering and a Normal selection strategy run
  on~\eqref{eq:KS} or~\eqref{eq:SMM}. As (SM-$\Cm$) contains (KS) directly
  into the system, computing a GB on~(SM-$\Cm$) will also compute all
  equations that would be computed by~(KS). On the other hand, when
  computing a GB for~(KS), the algorithm will produce all
  equations~(SM-$\Cm$) by multiplying by monomials in $\Cm$, hence we can
  expect many syzygies during a GB computation on~(SM-$\Cm$). 

  This encourages to compute with (SM-$\Cm$), but to look only at multiple
  of the equations by the $x_i$'s variables, which is the strategy
  proposed in~\cite{BBCGPSTV20}. Adding to this the change of variable
  that consider any minor of $\Cm$ as a variable removes the hardness
  of computing with high degree polynomials (as the new variables have
  degree 1 instead of a polynomial of degree $d$ with $d!$
  coefficients for the minor).
\end{remark}

\section{Complexity of solving superdetermined systems}
\begin{figure}
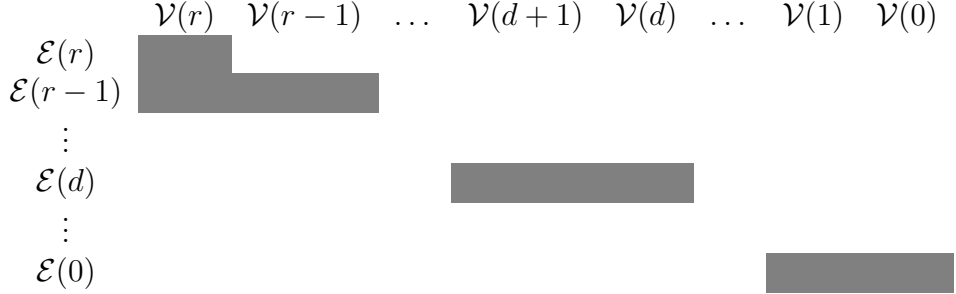

  \centering
  \begin{tabular}{*{9}{c}}
    & $\mathcal V({r})$ & $\mathcal V({r-1})$& $\dots$ & $\mathcal V({d+1}) $  & $\mathcal V({d})$ & $\dots $ &$\mathcal V({1})$ &  $\mathcal V({0})$\\
    $\mathcal E({r})$    &\cellcolor{gray} \\
    $\mathcal E({r-1})$ & \multicolumn{2}{c}{\cellcolor{gray}} \\
    \vdots & \\
    $\mathcal E({d})$ &  &&&\multicolumn{2}{c}{\cellcolor{gray}} \\
    \vdots & \\
    $\mathcal E({0})$& & &  &&&&\multicolumn{2}{c}{\cellcolor{gray}} \\ 
  \end{tabular}
  \caption{Shape of the Macaulay matrix associated to~\cref{eq:tout}. The columns correspond to equations in $\mathcal V(d)$, the rows to equations $\mathcal E(d)$. Gray cells correspond to non-zero part of the matrix.}
  \label{fig:shape}
\end{figure}

Superdetermined MinRank instances are defined in~\cite{VBCPS19} as MinRank instances where $K < rm$.
In the light of the previous section, it is now clear
that~\cite{VBCPS19} considers  for any $0\le d \le r$  the equations
\begin{align}
  \label{eq:tout}
  \mathcal E(d)\eqdef&\left\lbrace E_{J,T,\ell}\eqdef {\ev_{\ell}} \Mm_{\xv}  \begin{pmatrix}
    \Im_{n-r}\\{\Cm}
  \end{pmatrix} \trsp{\Vm_J(\Cm_{T,*})} : \substack{\forall J\subset\lbrace 1..n-r\rbrace, \#J=d+1,\\ \forall T\subset\lbrace 1..r\rbrace, \#T=d,\\\forall \ell\in\lbrace 1..m\rbrace}\right\rbrace.
\end{align}
and search for linear combination that will produce degree falls. We can rewrite the equations
\begin{align*}
  E_{J,T,\ell} &=  \minor{
  \begin{pmatrix}
\rv_\ell\\{\Cm} \; \Im_r
\end{pmatrix}}{*,T'} \text{ with } T'=J \cup (\lbrace n-r+1..n\rbrace\setminus (T+n-r))\subset\lbrace 1..n\rbrace\\
               &= \sum_{i=1}^K \sum_{j\in J} (\Mm_{i}^{(1)})_{\ell,j}x_i \minor{\Cm}{T,J\setminus \lbrace j\rbrace} + \sum_{i=1}^K \sum_{s\notin T} (\Mm_{i}^{(2)})_{\ell,s} x_i \minor{\Cm}{T\cup\lbrace s\rbrace,J}.
\end{align*}
We have a total of
$\sum_{d=0}^r m\binom{n-r}{d+1}\binom{r}{d}=m\binom{n}{r+1}$ equations
in $\sum_{d=0}^r K\binom{n-r}{d}\binom{r}{d}=K\binom{n}{r}$ variables
described by
\begin{align*}
  \mathcal V(d)\eqdef&\lbrace x_i
\minor{\Cm}{T,J}\rbrace_{i=1..K,\#J=d,\#T=d}, \quad \mathcal V(r+1)=\emptyset.
\end{align*}
The system can be solved by linearization by constructing the
associated Macaulay matrix: its rows are indexed by $J, T, \ell$ (with
$\#J=d+1$, $\#T=d$) and its columns by $i,J',T'$ (with $\#J'=d=\#T'$),
and the coefficient in row $(J,T,\ell)$ and column $(i,J',T')$
corresponds to the coefficient of $E_{J,T,\ell}$ in the monomial
$x_i\minor{\Cm}{T',J'}$. We can sort the columns by decreasing degree,
{\em i.e.} consider first monomials in $\mathcal V(r)$, up to
$\mathcal V(0)$ that are the $K$ variables $x_i$, see~\cref{fig:shape}. Then finding linear
combination of the equations that produce degree falls can be done
by computing the echelon form of the Macaulay matrix. For a set of rows
in $\mathcal E(d)$, we have $m\binom{n-r}{d+1}\binom{r}{d}$ equations
in $K\binom{n-r}{d+1}\binom{r}{r+1}+K\binom{n-r}{d}\binom{r}{d}$
monomials, and we get generically a degree fall under the condition
\begin{align*}
  m\binom{n-r}{d+1}\binom{r}{d} \geqslant K\binom{n-r}{d+1}\binom{r}{d+1}
\end{align*}
which is Corollary 5 in~\cite{VBCPS19}, and the first part of the
Macaulay matrix with columns in $\mathcal V(d+1)$ is, up to a good
choice of the ordering of rows and columns, a block of diagonal
matrices $\Bm_{J}$ as described in~\cite{VBCPS19}.

The best complexity estimates comes from the~\eqref{eq:SMMcT}
modeling, when considering the minors $\minor{\Cm}{T,J}$ as new
variables.  \cref{eq:tout} contains $m\binom{n}{r+1}$ equations in $K$
variables $x_i$ and $\binom{n}{r}$ variables that are minors of
$(-{\Cm} \; \Im_r)$. Hence the system can be solved whenever
$m\binom{n}{r+1}\geqslant K\binom{n}{r}$ by linearization, {\em i.e.}
$m(n-r)\geqslant K(r+1)$. After linearization, we get with overwhelming
probability $\#\mathcal V(0)-1=m-1$ linear equations in the $x_i$'s
only.

As always, it is possible to improve computation by puncturing the
matrix $\Mm_{\xv}$ (taking only sufficiently many columns so that we
keep an overdetermined system), or by hybrid approach (performing an
exhaustive search on some columns of $\Cm$, at the cost of $q^{ar}$
operation in $\mathbb{F}_{q}$ for $a$ columns). It is also possible, as
in~\cite[Eq. (23) p. 19]{BBCGPSTV20}, to compute equations at higher
degree $b$ in $x_i$. For instance, at $b=2$ we can multiply all
equations in (SM) by $x_i$'s variables, and we get for each set
$\mathcal E(d)$ of equations:
\begin{align}
  m\binom{n}{r+1}K-\binom{n}{r+2}\binom{m+1}{2} & \text{ equations, } & 
  \binom{n}{r}\binom{K+1}{2} & \text{ monomials}.
\end{align}

For instance,~\cref{tab:n10} compares the (SM) system with previous
results from \cite{VBCPS19}. For $r=5$, the ratio between equations
and monomials in~\eqref{eq:SMMcT} is smaller than 1, so that we cannot
expect to solve by linearization directly. Computing at $b=2$ would
produce 14400 equations in 13860 variables of degree less than or
equal to 7 in $(\xv,\Cm)$. Note that the last entry for $r=6$ would
theoretically require to go up to $b=5$ with matrices of size
427350. 

\begin{table}
  \centering
  \begin{tabular}{|c|c|c|c|c|c|c|c|}
  \hline
  $m$& $n$ & $K$ &$r$& $\frac{m(n-r)}{K(r+1)}$ & $n_{eq}$ & $n_{vars}$& $n_{rows}$ in~\cite{VBCPS19}\\\hline
  10 & 10 & 10 & 2  & $2.6$  & 1200 & 450& 1530\\
  10 & 5 & 10 & 2  & $1$  & 100 & 100& \\\hline
  10 & 10 & 10 & 3  & $1.75$  & 2100 & 1200& 20240\\
  10 & 7 & 10 & 3  & $1$  & 350 & 350& \\\hline
  10 & 10 & 10 & 4  & $1.2$  & 2520 & 2100 & 38586\\
  10 & 9 & 10 & 4  & $1$  & 1260 & 1260 & \\\hline  
  10 & 10 & 10 & 5  & {\bf 0.8}  & 2100 & 2520 & 341495\\\hline  
\end{tabular}
\caption{Size of matrices on~\eqref{eq:SMMcT} for a MinRank instance
  with $K=10$ matrices of size $m\times n$, for various $r$. $n$ can
  be decreased by puncturing the matrices to get a speedup. The
  results have been verified experimentally on random instances.}
\label{tab:n10}
\end{table}

However,~\cite{VBCPS19} suggest that we can have a closer look at the
shape of the equations and maybe find a better complexity for some
very overdetermined instances.

Hence, for a fixed $d\in\lbrace 0..r\rbrace$, the set $\mathcal E(d)$ contains
$m\binom{n-r}{d+1}\binom{r}{d}$ equations with
$K\binom{n-r}{d}\binom{r}{d}$ monomials $\mathcal V(d)$ of bidegree
$(1,d)$ and $K\binom{n-r}{d+1}\binom{r}{d+1}$ variables
$\mathcal V(d+1)$ of bidegree $(1,d+1)$. In~\cite{VBCPS19}, the
authors determine the first degree fall in KS by looking for the
smallest $d$ for which we have more equations in $\mathcal E(d)$ than
variables in $\mathcal V(d+1)$. This produces
$\Rank(\mathcal E(d))-\#\mathcal V(d+1)$ degree fall, but it is not
clear how to end the computation. If there is more equations than
variables, {\em i.e.}
\begin{align*}
  m\binom{n-r}{d+1}\binom{r}{d}\geqslant K\binom{n-r}{d+1}\binom{r}{d+1}+K\binom{n-r}{d}\binom{r}{d} -1
\end{align*}
then with overwhelming probability, the linear system is full rank
(its kernel has dimension 1 as the system is homogeneous in $\xv$) and
a non-zero element in the kernel of the Macaulay matrix gives a value
for each variable $x_i\minor{\Cm}{T,J}$. It is then straightforward to
deduce $x_i/x_{i_0}$ from two values $x_i\minor{\Cm}{T,J}$ and
$x_{i_0}\minor{\Cm}{T,J}$. On the other hand, if 
\begin{align*}
	K\binom{n-r}{d+1}\binom{r}{d+1}& \le   m\binom{n-r}{d+1}\binom{r}{d}\\
	\text{and }\qquad m\binom{n-r}{d+1}\binom{r}{d}&< K\binom{n-r}{d+1}\binom{r}{d+1}+K\binom{n-r}{d}\binom{r}{d} -1
\end{align*}
then it is necessary to add new equations to end the computation. It
can be done by consider equations of ``higher degree''. If each minor
of $\Cm$ is taken as a variable, it doesn't add a computational
burden. We can solve as soon as we can get sufficiently many blocks of
equations $\mathcal E(d)$ such that we get more equations than
columns. Experimental results are presented~\cref{tab:n12} on the same
parameters as~\cite[Table 2]{VBCPS19}. For instance for a MinRank
problem with $12\times 12$ matrices and a target rank $r=4$, the
authors in~\cite{VBCPS19} solve at degree $d=4$ in $58 s$, whereas it
is more interesting to consider equations in $\mathcal E(3..4)$ that
have degree up to $5$, without considering equations of degree
2..3. Note that if we puncture too much the matrices, for instance by
taking only $n=8$ columns, then we do not have any more an
overdetermined SM system, and solving it now require to produce more
equations, for instance by considering $b=2$. In this case, we get a
system of $5880 $ equations in $5460$ and we can solve, but this will
be more costly than solving with $n=9$.
\begin{table}
  \centering
\begin{tabular}{*{6}{|c}|*{6}{|c}|}
  \hline
  $r$& $\kappa$ & $d$ &  size & time &\cite{VBCPS19} \\\hline
  4 & 8 & $0..4$ & $9504 \times 5940$ & 5.6 s & 58 s\\
     & & $3..4$ & $4032\times 3528$ & 2.4 s & \\
     & 7 & $0..4$ & $5544\times 3960$ & 2.1 s & 38 s \\
     & 6 & $0..4$ & $3024 \times 2520 $ & 0.74 s & 21 s\\
  & & $2..4$ & $2232 \times 2220$ & 0.52 s & \\
     & 5 & $0..4$ & $1512 \times 1512 $ & 0.23 s & 13 s\\
  \hline
  5 & 7 & $0..5$ &  $11088 \times 9504$ & 11 s & 756 s\\
     & & $2..4$ & $9660 \times 9072$ & 9.6 s & \\
     & 6 & $0..5$ & $5544 \times 5544 $ & 3.1 s & 367s \\ \hline
\end{tabular}
\caption{Experimental size of matrices on SM for a MinRank instance
  with $K=12$ matrices of size $12\times 12$, for various $r$. It is
  possible to puncture the codes, by considering only $n=\kappa+r$
  columns of the matrices. We consider only systems for which SM
  solves at $b=1$. The second row gives the size of a submatrix of
  blocks $\mathcal E(d)$ for some $d$ that solves the problem faster.}
  \label{tab:n12}
\end{table}

\begin{remark}
  There is an asymetry between $m$ and $n$ in the modelings. It is
  always possible to exchange $m$ and $n$ by considering the transpose
  of the matrices, but it is not clear in general which problem will
  be easier ($m>n$ or $m<n$). For instance, for $K=10$, $r=2$ we can
  have the following behaviors: for $m=6, n=7$ the Macaulay matrix up
  to $d=r$ has size $210\times 210$, whereas for $m=7$, $n=6$ it is
  not possible to solve at $b=1$ (the Macaulay matrix has dimension
  $140\times 150$). We need to go to $b=2$ and solve a matrix of size
  $980\times 825$.  On the contrary, for $m=10$, $n=6$, 
  the Macaulay matrix has dimension $160\times 140$ ($d=1..2$),
  whereas for $m=6$, $n=10$, the Macaulay matrix for $d=r$ has size
  $336\times 280$.

  However, as the number of equations is a multiple of $m$, the best solution is often with $m\ge n$.
\end{remark}

\section{Application to DAGS}%
\label{sec:application_on_dags} 
DAGS scheme~\cite{BBBCDGGHKNNPR17} is a key encapsulation mechanism (KEM) based on
quasi-dyadic alternant codes that was submitted to the first round of
the NIST standardization process for a quantum resistant public key
algorithm. It suffered from an algebraic attack~\cite{BC18} that
efficiently recovers the private key, and was improved
in~\cite{BBCO19}. Here, we show that the DAGS algebraic modeling is in
fact a MinRank problem. However, the previous complexity results do
not apply, as those MinRank instances have a structure, that can be
used to understand more precisely the complexity.

\subsection{Principle of the attack}%
\label{sub:principle_of_the_attack}

We recall some elements of the scheme.
DAGS is based on the McEliece scheme and uses Quasi-Dyadic Generalized Srivastava codes, which are a subfamily of alternant codes.
The structure of such codes is what allowed DAGS to be attacked~\cite{BC18,BBCO19}.

The idea of the key-recovery attack leading to the modeling presented here is 
to find a subcode of the public code. The attack was proposed in two versions: a combinatorial one that uses brute force to find the subcode, and an algebraic one that relies on solving a polynomial system. The complexity of the combinatorial version is easy to compute, however the numbers of calculations remains too high to be done in practice. On the contrary, the algebraic attack is more efficient but its complexity is harder to estimate. 

We focus on the second version and explain the principle. 
We begin by computing the invariant subcode of the public code of the scheme.
Then, we search for a subcode of this invariant code by solving a bilinear system built from public parts of the scheme.
Finally, we can recover the support and multiplier of the original alternant code.

In the next subsection, we explain how the system we want to solve is built.

\subsection{Original Modeling}%
\label{sub:original_modeling}

Let $\Cpub$ be the DAGS public code, $\Hpub$ be the public key of the scheme, which is a parity-check matrix of $\Cpub$, and let $\Gpub$ be its generator matrix.

We refer to~\cite[Chap. 12]{MS86} for the definition of alternant codes. DAGS codes are quasi-dyadic alternant codes over $\mathbb{F}_{q^m}$, with $q$ a power of $2$ and $m=2$.
To build the system we need to understand the construction of quasi-dyadic alternant codes, that are alternant codes for which the support $\xv$ and the multiplier $\yv$ have a particular structure. 
\begin{definition}
  Let $\gamma \geqslant 1$ and $n=2^\gamma n_0$.
  The support $\xv\in\mathbb{F}_{q^m}^n$ of a quasi-dyadic alternant code of order $2^\gamma$  is constructed from $(b_1,\dots,b_\gamma)\in \mathbb{F}_{q^m}^\gamma$ that are linearly independent over $\mathbb{F}_2$, and $\tauv=(\tau_1,\dots,\tau_{n_0})\in \mathbb{F}_{q^m}^{n_0}$ as
  \begin{align*}
    \xv&\eqdef \tau \otimes \onev_{2^\gamma}+\onev_{n_0}\otimes \gv,
  \end{align*}
  where $\gv\eqdef (g)_{g\in\mathbb G}$ is a vector of all $2^\gamma$ elements of the group $\mathbb G = \langle b_1,\dots,b_\gamma\rangle_{\mathbb{F}_2}$ which is the vector space generated by the elements $(b_i)$ over $\mathbb{F}_2$.
  
  The elements $\tau_i$ are randomy drawn from $\mathbb{F}_{q^m}$ such that the cosets $\tau_i+\mathbb G$ are pairwise disjoint.
\end{definition}
For instance for $\gamma=2$, we can choose $\gv = (0,b_1,b_2,b_1+b_2) = b_1(0,1,0,1)+b_2(0,0,1,1)$. For $\gamma=3$ we take $\gv = (0,b_1,b_2,b_1+b_2,b_3,b_1+b_3,b_2+b_3,b_1+b_2+b_3) = b_1(0,1,0,1,0,1,0,1)+b_2(0,0,1,1,0,0,1,1) + b_3(0,0,0,0,1,1,1,1)$. In general, one possible order for $\gv$ is given by $  \gv = \sum_{i=1}^\gamma b_i \ev_i$ where
\begin{align*}
  \ev_i &\eqdef (\zerom_{2^{i-1}},\onev_{2^{i-1}}, \zerom_{2^{i-1}},\onev_{2^{i-1}},\dots) = \onev_{2^{\gamma-i}}\otimes (0,1)\otimes \onev_{2^{i-1}}. 
\end{align*}

The group $\mathbb G$ acts by translation on $\mathbb{F}_{q^m}$, and its action induces a permutation of the code $\Cpub$. This is what allows the DAGS system to have reduced public keys: the public matrix $\Gpub$ is formed by blocks of size $2^\gamma$ where each row of the block is deduced from the first row by one of the permutation induced by $\mathbb G$.

The attack in \cite{BC18} introduces the \emph{invariant subcode} $\Cpub^{\mathbb G}$ with respect to $\mathbb G$ of $\Cpub$, which is defined as
\begin{definition}\label{def:invariant_code}
  The \emph{invariant code} of $\Cpub$ is defined by: 
  \begin{equation*}
    \Cpub^{\mathbb G} = \left\{ \cv \in \Cpub | \forall (i,j)\in\lbrace 0..n_0-1\rbrace \times \lbrace 1..2^\gamma\rbrace, c_{i2^\gamma+j}=c_{i2^\gamma+1}  \right\}.
  \end{equation*}
\end{definition}
The invariant subcode has dimension $k_0=k/2^\gamma$ where $k$ is the
dimension of $\Cpub$. Its generator matrix $\Ginv$ is easy to compute
from $\Gpub$: each block of $2^\gamma$ rows of $\Gpub$ gives one row of $\Ginv$ by summation. The entries of $\Gpub$ are then repeated by blocks of size $2^\gamma$, so that we can define a matrix $\tilde{\Gm}\in\mathbb{F}_{q^m}^{k_0\times n_0}$ satisfying $\Ginv = \tilde{\Gm}\otimes\onev_{2^\gamma}$.

We introduce the component-wise product called Schur product:
\begin{definition}
        The \emph{Schur product} of two codes $\code{A}$ and $\code{B} \subseteq \mathbb{F}_{q}^{n}$ corresponds to the code generated by all the component-wise products of one codeword from $\code{A}$ and one codeword of $\code{B}$:
        \begin{equation*}
                \code{A} \star \code{B} = \left< \av \star \bv \mid \av \in \code{A}, \bv \in \code{B} \right>_{\mathbb{F}_{q}}
        \end{equation*}
\end{definition}
The attack in~\cite{BC18} amounts to find $\code{D}$, an unknown subcode of $\Cpub^{\mathbb G}$ such that $\xv$ is orthogonal to ${ \code{D} \star \dual{\Cpub} }$.
This leads to following system with 2 unknowns, $\code{D}$ and $\xv$:
\begin{equation} \label{eq:dagssystem_x}
	\Gm_{\code{D} \star \dual{\Cpub}} \cdot \trsp{\xv} = 0
\end{equation}
Algebraically,  a generator matrix for $\code{D} \star \dual{\Cpub}$ can be written with high probability as
\begin{equation}
 \left( \begin{pmatrix} \Im_{k_{0}-c} &  \Um \end{pmatrix} \cdot \Gm_{inv} \right) \star \Hm_{pub} 
\end{equation}
with $c$ the codimension of $\code{D}$ in the invariant subcode $\Cpub^{\mathbb G}$. 
If we can not express the system like that, we just need to take another generator matrix for the invariant subcode of $\Cpub$.
This finally leads to the original modeling:
\begin{align}
   \label{eq:dagsori}
\left(   \begin{pmatrix}
     \Im_{k_0-c} & \Um
   \end{pmatrix}
                     \Gm_{inv}\star \Hm_{pub}\right)  \trsp{\xv} = 0
\end{align}
where $\Um$ is a matrix of unknowns of size $(k_0-c)\times c$, 
$\Gm_{inv} = \tilde{\Gm}\otimes \onev_{2^\gamma}$ and
$\tilde{\Gm} $ is a public invariant matrix of size
$k_0\times n_0$, $\Hm_{pub}$
 is the public parity-check matrix, and
$\xv = \tauv\otimes \onev_{2^\gamma} + 
\sum_{i=1}^\gamma b_i \onev_{n_0}\otimes \ev_i \in
\mathbb{F}_{q^m}^{n}$ is a vector of unknowns
$\tauv=(\tau_1,\dots,\tau_{n_0})$ and $(b_1,\dots,b_{\gamma})$.
\begin{remark}
  As explained in~\cite{BC18}, any affine map $\xv\to a\xv+b$ for $a\in\mathbb{F}_{q^m}^*, b\in \mathbb{F}_{q^m}$ preserves the quasi-dyadic structure of the code, and leaves the code invariant, so that it is always possible to search among all possible $\xv$ for the ones that satisfy $b_1=1$ and $\tau_{n_0}=0$.
  Moreover, 
  the vector $\xv^q$, hence $\tr(\xv) \eqdef \xv + \xv^q$ are also
  solution of the system~\eqref{eq:dagsori}, so that $\tr(b_2)^{-1}\tr(\xv)$ is a solution with $\tau_{n_0}=0$, $b_1=0$ and $b_2=1$ (as $\tr(a)=0$ for $a\in\mathbb{F}_{q}$ when $m=2$).
\end{remark}
\begin{remark}
  As explained in~\cite{BBCO19},  there is a lot of redundancy among the equations. 
  We avoid that  by considering  only one out of every $2^{\gamma}$ rows in $\Hpub$.
\end{remark}

\def\kmoinsa{k_0} \def\nmoinsa{n_0}
\def\nmoinsk{n_0-k_0}

\subsection{Modeling Update}%
\label{sub:modeling_update}

A simple (but fastidious, see~\cref{sec:appendix}) computation allows to write the system as a
MinRank instance with matrices  of size  $(\nmoinsk) \times \kmoinsa$, when $\tilde{\Gm}=(\Im_{k_0} \Gm)$ is taken in systematic form:
\begin{align}\label{eq:dagsminrank}
	\left(\sum_{i=1}^{\kmoinsa}\tau_i\Mm_{i}{}+\sum_{j=1}^{\nmoinsk}\tau_{j+\kmoinsa}\Mm_{j+\kmoinsa}{} +\sum_{b=2}^\gamma b_i\Hm_i \right)
  \begin{pmatrix}
    \Im_{\kmoinsa-c}\\\trsp{\Um}
  \end{pmatrix} = \zerom
  \\
  \text{ with } \Mm_{i}{} =
   \begin{pmatrix}
     \zerom_{i-1} & \trsp{(\Gm_{\{i\},*})} & \zerom_{\kmoinsa-i}
   \end{pmatrix}\;  \forall 1\le i \le \kmoinsa \notag{}\\
  \Mm_{j+\kmoinsa}{} =
  \begin{pmatrix}
    \zerom_{j-1}\\
    \trsp{(\Gm_{*,\{j\}})}\\
    \zerom_{\nmoinsk-j}
  \end{pmatrix} \forall 1\le j \le \nmoinsk \notag{}\\
  \Hm_i = \Hpub(\Im_{\nmoinsa}\otimes \trsp{\ev_i})_{*,\lbrace 1..k_0\rbrace}\; \forall 2\le
  i\le \gamma\notag{}
\end{align}

It is clear that the matrices $\Mm_{i}{}$ from DAGS instances are not
random, and in practice we have more degree falls than expected. On
the other hand, the part concerning the variables $b_i$ with matrices
$\Hm_i$ seems to behave like a random system. Note also that
experimentally we find that the system always produces 3
solutions. However, this is small enough to be able to recover the
good one from the kernel of the Macaulay matrix, as only one belong to
the finite field $\mathbb{F}_{q}$.

\begin{proposition}
  For the DAGS modeling, the 
  Macaulay matrix associated to the set of equations $\mathcal E(d)$  
has  size $N_{rows}\times N_{cols} = (n_0-k_0)\binom{\kmoinsa-c}{d+1}\binom{c}{d} \times (n_0-k_0-1+c+\gamma-1)\binom{\kmoinsa-c}{d+1}\binom{c}{d+1} $, but its rank is
  \begin{align*}
    \Rank(\mathcal E(d)) &= \min\left( N_{rows}, \binom{\kmoinsa-c}{d+1}\left((n_0-k_0)\binom{c-1}{d}+\binom{c}{d+1}d \right)\right)
  \end{align*}
\end{proposition}

Note that it is always possible to use shortened codes on $a_0$ positions, that amounts to consider codes with parameters $(n_0-a_0, k_0-a_0)$.

The first sets of parameters were given in the specifications of the scheme. 
They are shown in~Table\ref{tab:dags_original_parameters}. 
Experimental results in~\cite{BBCO19} give a solution of the system
DAGS\_3 in degree 4 with linear algebra on a matrix of size
$725,895 \times 671,071$. It is improved by shortening the system up
to $k_0-a_0-c=4$ with a matrix of size $103,973 \times 97,980$ and a
computation lasting 70 seconds. All results presented here allows to
choose to shorten the system to $k_0-a_0-c=5$ instead of 4, as for 4
the system does not leads directly to linear equations, and it reduces
the computation to linear algebra on a matrix of size 2772 by 4284 that
last only few seconds.

\begin{table}
  \centering
  \begin{tabular}{*{10}{|c}|}
    \hline
    Security Level & $q$     & $n_0$ & $k_0$ & $\gamma$ & $c$ & $k_0-a_0-c$ & Matrix size        & Rank & Time   \\
    \hline
    DAGS\_1 (128)  & $2^{5}$ & $52$  & $26$  & $4$      & $4$ & 4           & $1456 \times 2520$ & 1322 & $3.5$s \\
    DAGS\_3 (192)  & $2^{6}$ & $38$  & $16$  & $4$      & $4$ & 5           & $2772\times 4284$  & 2540 & $8.8$s \\
    DAGS\_5 (256)  & $2^{6}$ & $33$  & $11$  & $2$      & $2$ & 3           & $220 \times 310$   & 194  & $0.0$s \\
    \hline
  \end{tabular}
  \caption{DAGS original sets of parameters}
  \label{tab:dags_original_parameters}
\end{table}

\section*{Conclusion}
We have presented the link between the different modelings for the
MinRank problem. This allows a more accurate understanding of the best
strategy to solve  MinRank instances.

We have shown that superdetermined MinRank instances are instances for which (SM) solves at $b=1$, and that the maximal degree in the computations is not the best parameter to use to optimize the computation.
 
We have also presented the DAGS attack as a particular superdetermined MinRank one,
and how the accurate study of the involved matrices allows to find the
best strategy.

\subsubsection{Acknowledgements}
This work has been supported by  the French ANR project
CBCRYPT (ANR-17-CE39-0007).

\appendix
\section{Appendix}
\label{sec:appendix}
We want to reduce~\eqref{eq:dagsori} to a MinRank problem~\eqref{eq:dagsminrank}. We start from~\eqref{eq:dagsori}:
\begin{align*}
  \left(   \begin{pmatrix}
     \Im_{k_0-c} & \Um
   \end{pmatrix}
                     (\tilde{\Gm}\otimes \onev_{2^\gamma})\star \Hm_{pub}\right)  \trsp{\xv} = 0.
\end{align*}
Using the fact that $(\Am\star\Bm)\trsp{\av}=0$ is equivalent to $(\Am\star\av)\trsp{\Bm}=0$, that  $\Am\otimes \av = \Am(\Im\otimes \av)$ and $(\Am\Bm)\star\av=\Am(\Bm\star\av)$, it can be rewritten
\begin{align*}
   \begin{pmatrix}
     \Im_{k_0-c} & \Um
   \end{pmatrix}
                     \tilde{\Gm}((\Im_{n_0}\otimes \onev_{2^\gamma})\star \xv) \trsp{\Hm_{pub}} = 0
\end{align*}
Now we can use the relations
$(\Am\otimes \av)\star(\bv\otimes \xv) =
(\Am\star\bv)\otimes(\av\star\xv)$, $\tauv\star \Im=\Diag(\tauv)$ and
$\Am\otimes\av = \Am(\Im\otimes\av)$ to simplify
\begin{align*}
  (\Im_{n_0}\otimes \onev_{2^\gamma})\star \xv &=   (\Im_{n_0}\otimes \onev_{2^\gamma})\star(\tauv\otimes \onev_{2^\gamma} + \sum_{i=1}^\gamma b_i \onev_{n_0}\otimes \ev_i )\\
                                               &= (\tauv\star\Im_{n_0})\otimes \onev_{2^\gamma} + \sum_{i=1}^\gamma b_i (\Im_{n_0}\otimes \ev_i)\\
  &= \Diag(\tauv)(\Im_{n_0}\otimes \onev_{2^\gamma}) + \sum_{i=1}^\gamma b_i (\Im_{n_0}\otimes \ev_i)
\end{align*}
We can now define $  \tilde{\Hm_i} = \Hpub(\Im_{\nmoinsa}\otimes \trsp{\ev_i})$ and $\tilde{\Hm}=\Hpub(\Im_{n_0}\otimes \trsp{\onev_{2^\gamma}})$ and we get the system
\begin{align*}
   \begin{pmatrix}
     \Im_{k_0-c} & \Um
   \end{pmatrix}
                     \tilde{\Gm}(\Diag(\tauv) \trsp{\tilde{\Hm}} + \sum_{i=1}^\gamma b_i \trsp{\tilde{\Hm_i}}) = 0,\\
  (\tilde{\Hm}\Diag(\tauv)\trsp{\tilde{\Gm}} + \sum_{i=1}^\gamma b_i {\tilde{\Hm_i}}  \trsp{\tilde{\Gm}}) \begin{pmatrix}
     \Im_{k_0-c} \\ \trsp{\Um}
   \end{pmatrix}
= 0
\end{align*}
We now simplify the products using the remark  that $\tilde{\Hm}$ is the parity-check matrix corresponding to $\tilde{\Gm}=(\Im_{k_0} \; \Gm)$: $\tilde{\Hm}=(\trsp{\Gm} \; \Im_{n_0-k_0})$, and that $\tilde{\Hm_i}=(\Hm_i \; \zerom_{n_0-k_0})$ contains columns of zeros on the last $n_0-k_0$ positions. This gives~\eqref{eq:dagsminrank}.

\end{document}